\documentclass[a4paper,11pt]{article}
\usepackage{pos}

\usepackage{amsmath,amsfonts}

\newcommand{\be}{\begin{equation}}
\newcommand{\ee}{\end{equation}}
\newcommand{\bes}{\begin{equation*}}
\newcommand{\ees}{\end{equation*}}
\newcommand{\bdis}{\begin{displaymath}}
\newcommand{\edis}{\end{displaymath}}
\newcommand{\bga}{\begin{equation}\begin{gathered}}
\newcommand{\ega}{\end{gathered}\end{equation}}
\newcommand{\bgas}{\begin{equation*}\begin{gathered}}
\newcommand{\egas}{\end{gathered}\end{equation*}}
\newcommand{\Tr}{\mathop{\mathrm{Tr}}\nolimits}

\title{QCD topology and axion's properties from Wilson twisted mass lattice simulations}

\author[a,b]{A.Yu.~Kotov}
\author[c]{M.P.~Lombardo}
\author*[d]{A.~Trunin}

\affiliation[a]{J\"ulich Supercomputing Centre, Forschungszentrum J\"ulich, D-52428 J\"ulich, Germany}
\affiliation[b]{Bogoliubov Laboratory of Theoretical Physics, Joint Institute for Nuclear Research, Dubna, 141980 Russia}
\affiliation[c]{INFN, Sezione di Firenze, 50019 Sesto Fiorentino (FI), Italy}
\affiliation[d]{Samara National Research University, Samara, 443086 Russia}

\emailAdd{a.kotov@fz-juelich.de}
\emailAdd{lombardo@fi.infn.it}
\emailAdd{amtrnn@gmail.com}

\abstract{We present the results on topological susceptibility and chiral observables in $N_f=2+1+1$ QCD for temperature range $120<T<600$ MeV. The lattice simulations are performed with Wilson twisted mass fermions at physical pion, strange and charm masses. 
In high-$T$ region $T\gtrsim 300$ MeV
the chiral observables are shown to follow leading order Griffith analyticity, 
and the topological susceptibility follows a power-law decay as in the instanton dilute gas models. The measured topological susceptibility is used to estimate the mass of QCD axion. The resulting axion mass constraints are in agreement with our previous studies at higher pion masses.}

\FullConference{%
 The 38th International Symposium on Lattice Field Theory, LATTICE2021
  26th-30th July, 2021
  Zoom/Gather@Massachusetts Institute of Technology
}

\begin{document}
\maketitle

\section{Introduction}
The topological aspects of QCD play a pivotal role in many theoretical problems. 
Prominent examples include the explanation of the $\eta'$ meson mass~\cite{Witten:1979vv,Veneziano:1979ec} and (possible) solution to the strong CP problem leading to the prediction of a new particle, the QCD axion~\cite{Peccei:1977hh,Weinberg:1977ma,Wilczek:1977pj}. This new particle is also considered as a promising candidate for Dark Matter constituent. Another wide topic of interest is the interplay between topology and various mechanisms of chiral and axial symmetry breaking/restoration in hot QCD~\cite{Gross:1980br,Ringwald:1999ze,Bottaro:2020dqh}.

Lattice simulations were first applied to the problem of axion properties in Ref.~\cite{Berkowitz:2015aua}. In particular, the axion mass can be extracted from lattice data on high-temperature topological susceptibility under certain assumptions about axion cosmological evolution (post-inflationary scenario). First results were obtained in~\cite{Berkowitz:2015aua} in quenched approximation, followed by numerous works with dynamical quarks~\cite{Bonati:2015vqz,Borsanyi:2016ksw,Petreczky:2016vrs,Bonati:2016tvi,Burger:2017xkz,Burger:2018fvb}. 

In this Proceeding we report the preliminary results of our ongoing project on simulation of finite-$T$ QCD with Wilson twisted mass fermions at the physical point. We extend our previous study on axions performed at higher than physical pion masses in~\cite{Burger:2017xkz,Burger:2018fvb}.
We calculate the temperature dependence of several chiral observables, including chiral condensate and susceptibility, and relate them to high-temperature topological susceptibility via QCD symmetry relations. Then, using the observed value of Dark Matter density as an input, we obtain the lower limit on (post-inflationary) axion mass.

\section{Lattice setup}
We perform simulations with $N_f=2+1+1$ Wilson twisted mass fermions tuned at maximal twist~\cite{Frezzotti:2003ni,Shindler:2007vp}.
The summary of our lattice ensembles are given in Table~\ref{tbl:summary}. Strange and charm quark masses are set to the physical values, and four different pion masses are available including the physical point. For lattice spacing and other parameters we rely on ETMC $T=0$ results~\cite{Alexandrou:2014sha,Alexandrou:2020okk}.
We employ fixed-scale approach for finite-$T$ simulations: for each ensemble the lattice spacing $a$ is fixed, and the temperature is varied by lattice size in temporal direction $L_t$. Thus we cover the temperature range approximately  $120 \lesssim T \lesssim 600$~MeV. 
Additional details on our lattice simulations can be found in~\cite{Burger:2018fvb,Kotov:2021rah}.

\begin{table}[thb]
\begin{center}
\begin{tabular}{c c c}
\hline
\hspace*{.2cm}Ensemble\hspace*{.2cm} & \hspace*{.2cm}$m_\pi$ [MeV]\hspace*{.2cm} & \hspace*{.2cm}$a$ [fm]\hspace*{.2cm} \\
\hline
M140  & 139(1) & 0.0801(4) \\
D210  & 213(9) & 0.0646(7) \\
A260 & 261(11) & 0.0936(13) \\
B260 & 256(12) & 0.0823(10) \\
A370 & 364(15) & 0.0936(13) \\
B370 & 372(17) & 0.0823(10)  \\
D370 & 369(15) & 0.0646(7)  \\
\hline
\end{tabular}

\caption{Parameters of $N_f=2+1+1$ lattice ensembles used for the analysis~\cite{Alexandrou:2014sha,Alexandrou:2020okk}.
\label{tbl:summary}
}
\end{center}
\end{table}

\section{Observables}
We consider the following chiral observables:
\begin{itemize}
\item Chiral condensate $\langle \bar{\psi}\psi\rangle=\langle \bar{u}u\rangle+\langle \bar{d}d\rangle=\dfrac{T}{V}\dfrac{\partial Z}{\partial m_l}=\dfrac{1}{L_t L_s^3}\langle\Tr M^{-1}\rangle$.
\item Chiral susceptiblity $\chi_L=\dfrac{\partial}{\partial m_l}\langle \bar{\psi}\psi\rangle=\chi_\text{disc}+\chi_\text{conn}$
consisting from connected and disconnected parts.
\item By combining chiral condensate $\langle \bar{\psi}\psi\rangle$ and its susceptibility $\chi_L$ we introduce the new observable
\begin{equation}
    \langle\bar\psi\psi\rangle_3 = \langle\bar\psi\psi\rangle - m_l\, \chi_L,
\label{eq:pbp3-def}
\end{equation}
which is free from linear additive renormalization as well as from linear correction to scaling. For additional details on $ \langle\bar\psi\psi\rangle_3$ and its properties we refer to~\cite{Kotov:2021rah,lat21_proc}.
\end{itemize}

In order to measure the topological susceptibility $\chi_\text{top}$ we employ its relation to the disconnected chiral susceptibility $\chi_\text{disc}$ 
via the QCD symmetry arguments~\cite{Kogut:1998rh,Bazavov:2012qja,Buchoff:2013nra}. In particular, the following continuum relation is valid:
\begin{equation}
\label{eq:chit-chi5}
\chi_\text{top}=\frac{\langle Q^2\rangle}{V}=m_l^2\,\chi_{5,\text{disc}},
\end{equation}
where $Q$ is topological charge, and $\chi_{5,\text{disc}}$ is disconnected pseudo-scalar susceptibility. The direct measurement of $\chi_{5,\text{disc}}$ on the lattice is difficult due to large fluctuations. Instead, we note that after the chiral transition $\chi_{5,\text{disc}}$ becomes equal to $\chi_\text{disc}$. Then, Eq.~\eqref{eq:chit-chi5} reads as
\begin{equation}
\label{eq:chit-pbp}
\chi_\text{top}(T\gtrsim T_c)=m_l^2\,\chi_\text{disc}=m_l^2\,\frac{V}{T}\left( \langle{(\bar\psi \psi)^2}\rangle_l - \langle{\bar\psi \psi}\rangle_l^2 \right)
\end{equation}
defining the topological susceptibility in high-$T$ region. Finally, we note that Eqs.~\eqref{eq:chit-chi5}--\eqref{eq:chit-pbp} are exact only in the continuum limit, meaning that  fine lattices should be used in order to avoid large artifacts.

\section{Results}
\begin{figure}[bt]
\begin{center}
\includegraphics{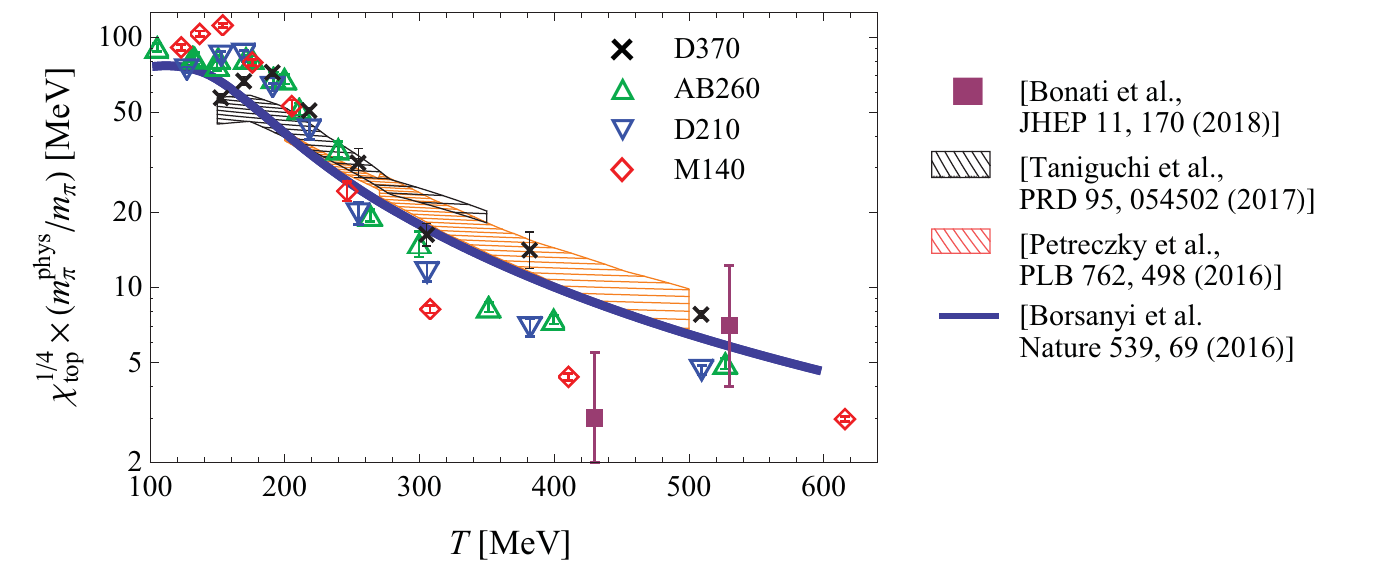}
\end{center}
\vspace*{-.3cm}
\caption{Topological susceptibility vs temperature obtained in this work and in Refs.~\cite{Bonati:2018blm,Taniguchi:2016tjc,Petreczky:2016vrs,Borsanyi:2016ksw,Burger:2018fvb}.  The results for non-physical pion masses are rescaled as $\chi_\text{top}\propto m_\pi^4$.
\label{fig:top_results}}
\end{figure}

We present the results on topological susceptibility measured according to Eq.~\eqref{eq:chit-pbp} at the physical pion mass in Fig.~\ref{fig:top_results}. We compare it with the results obtained in other lattice approaches~\cite{Bonati:2018blm,Taniguchi:2016tjc,Petreczky:2016vrs,Borsanyi:2016ksw} and also with our previous study at higher pion masses~\cite{Burger:2018fvb}. 
In order to set the common scale for comparison, the results from non-physical pion masses are rescaled according to $\chi_\text{top}\propto m_\pi^4$. Such behavior is predicted by dilute instanton gas model (DIGA) and can also be obtained from more general considerations based on the analyticity of chiral condensate in light quark mass~\cite{Burger:2018fvb}.
Fig.~\ref{fig:top_results} shows that different studies lead to similar results following the same trend, but still lacking  complete numerical agreement.

\begin{figure}[tb]
\vspace*{.2cm}
\begin{center}
\includegraphics{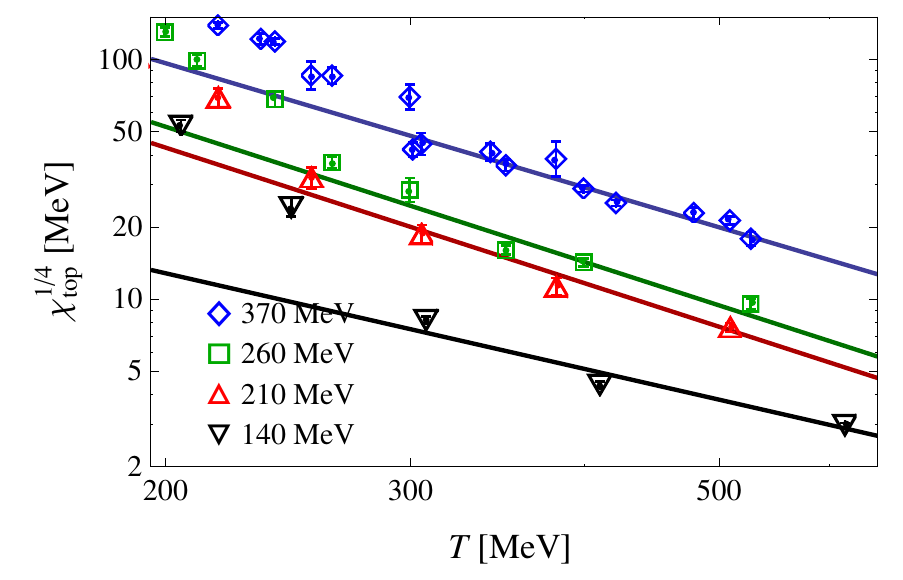}
\end{center}
\vspace*{-.3cm}
\caption{Fits of the topological susceptibility with power-law decay~\eqref{eq:chi_top-diga}.
All ensembles from Table~\ref{tbl:summary} corresponding to the same value of pion mass are treated equally.
\label{fig:top_fits}
}
\end{figure}

In order to obtain a simple analytical expression for topological susceptibility, we fit it in Fig.~\ref{fig:top_fits} with DIGA-inspired high-temperature behavior 
\be
\chi_\text{top}\simeq A\,T^{-d}.
\label{eq:chi_top-diga}
\ee
The data are well described by the power-law decay~\eqref{eq:chi_top-diga} in the region $T\gtrsim300$~MeV. 
For higher than physical pion masses the fits are performed over the combined data from all available ensembles (see Table~\ref{tbl:summary}).
The data show no apparent lattice spacing dependence suggesting that artifacts are small, as was also confirmed in~\cite{Burger:2018fvb} by more detailed analysis.

\begin{figure}[tb]
\begin{center}
\includegraphics[width=.6\linewidth]{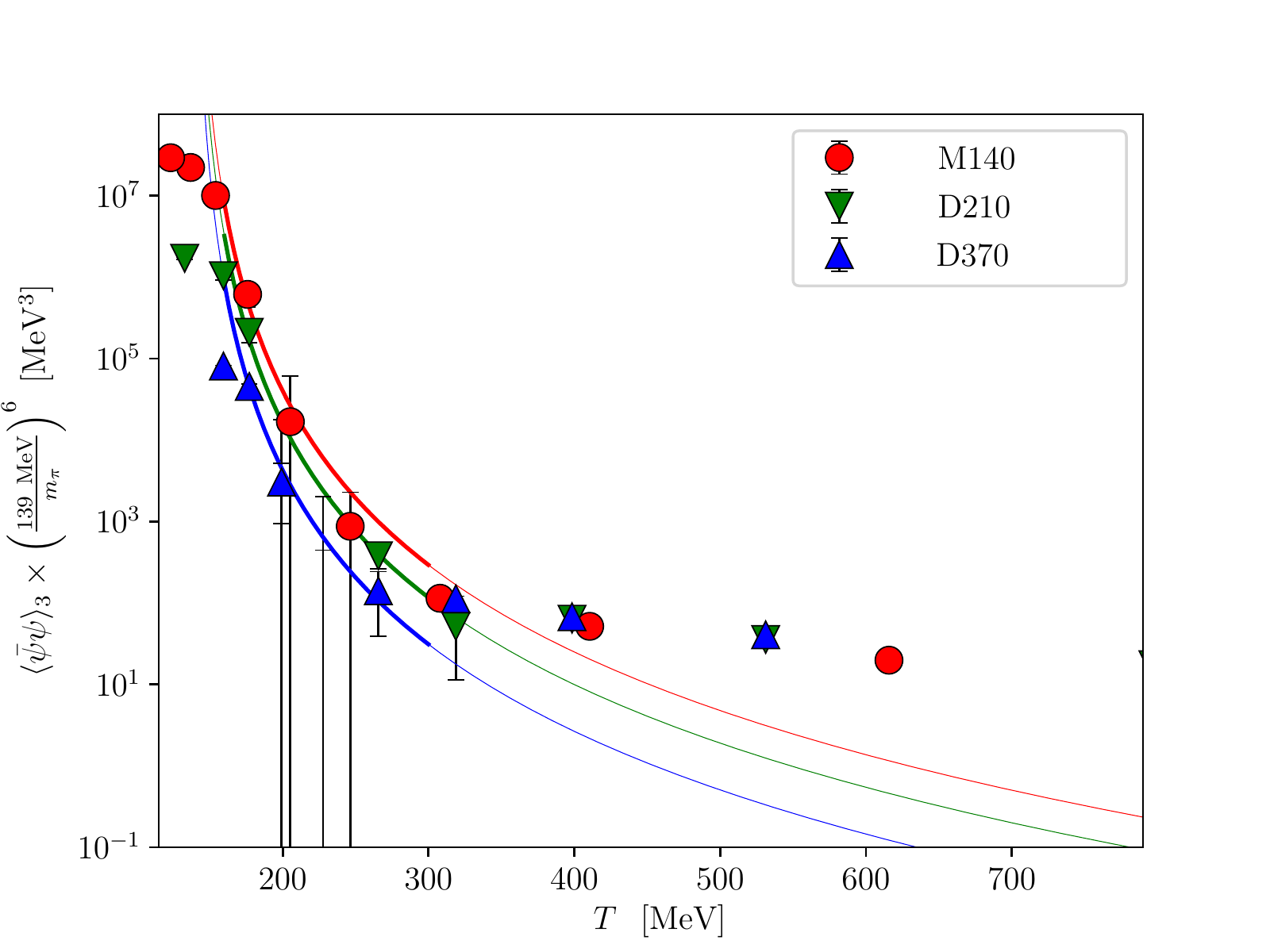}
\end{center}
\caption{$\langle\bar\psi\psi\rangle_3$ vs temperature, also fitted with the 3D $O(4)$ scaling behavior $\langle\bar{\psi}\psi\rangle_3\propto (T-T_0)^{-\gamma-2\beta\delta}$. For higher than physical pion masses the data are rescaled as $\langle\bar{\psi}\psi\rangle_3\propto m_\pi^6$.
\label{fig:pbp3}
}
\end{figure}

The temperature dependence of the $\langle\bar{\psi}\psi\rangle_3$~\eqref{eq:pbp3-def} is shown in Fig.~\ref{fig:pbp3}. First, we rescale with the leading order Griffith analyticity prediction $\langle\bar{\psi}\psi\rangle_3\propto m_\pi^6$. Then, we fit with the universal scaling behavior $\langle\bar{\psi}\psi\rangle_3\propto (T-T_0)^{-\gamma-2\beta\delta}$, where $T_0$ is fixed to the critical temperature $T_0=138$ MeV in the chiral limit~\cite{Kotov:2021rah,lat21_proc}. The critical exponents $\beta$, $\gamma$ and $\delta$ are fixed to represent 3D $O(4)$ universality class. 
As expected, the universal behavior sets in near the transition and remains up to  $T\simeq300$~MeV.
After that, rescaled data from different pion masses merge to the single curve, indicating simple Griffith analyticity behavior.
It is intriguing that this change of trend in $\langle\bar{\psi}\psi\rangle_3(T)$ coincides with the onset of DIGA-like behavior for the topological susceptibility $\chi_\text{top}(T)$ mentioned above, both occurring at approximately the same temperature $T\simeq300$~MeV.

Once we determined the temperature dependence of topological susceptibility~\eqref{eq:chi_top-diga} in high-$T$ region, we can use it to estimate the axion mass~\cite{Turner:1985si,Berkowitz:2015aua}:
\be
m_A(T)=\frac{\sqrt{\chi_\text{top}(T)}}{f_A}.
\label{eq:mA}
\ee
Since the exact value of the axion decay constant $f_A$ is unknown, we take relation~\eqref{eq:mA} at two moments of time (or, equivalently, temperatures) corresponding to the evolution of axions in the early Universe and to present day. The two time moments are connected by the axion equation of motion, allowing to obtain today's axion density $\Omega_A$ as a function of its mass.
Then, assuming that axions are responsible for the observed Dark Matter density $\Omega_\text{DM}$, the axion mass can finally be extracted.
For detailed derivation we refer to the original works~\cite{Turner:1985si,Berkowitz:2015aua} (see also the review~\cite{Lombardo:2020bvn} and references therein). In particular, we use the result of Ref.~\cite{Burger:2018fvb}
\be
\Omega_A=F(A,d,\ldots)\,m_A^{-\frac{3.053+d/2}{2.027+d/2}},
\label{eq:omega}
\ee
where $F$ is a function of topological susceptibility parameters~\eqref{eq:chi_top-diga} (amplitude~$A$ and power decay constant~$d$) and of  relevant cosmological constants.

\begin{figure}[tb]
\begin{center}
\includegraphics{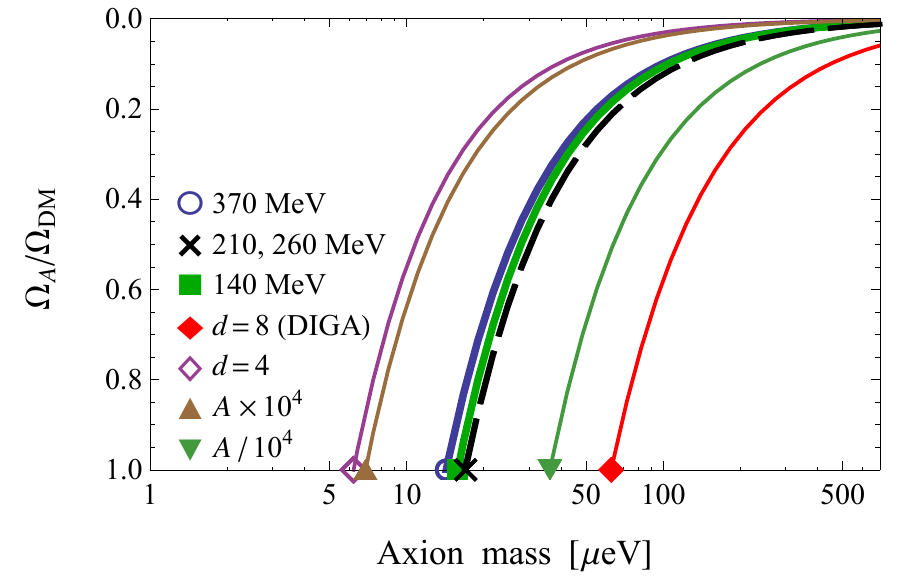}
\end{center}
\caption{The axion fraction in Dark Matter vs the axion mass.
For physical pion ensemble the parameters $A$ and $d$ are varied as indicated in the legend.
\label{fig:omega}
}
\end{figure}

We plot the result~\eqref{eq:omega}, using the parameters extracted from our fits, in Fig.~\ref{fig:omega}. For physical pion ensemble we also explore the limiting cases by increasing or decreasing the amplitude $A$ by factor $10^4$ and setting the decay constant  $d=8$ and $d=4$ corresponding to pure DIGA prediction and to very slow decay of topological susceptibility, respectively.
The actual fraction of axion density in $\Omega_\text{DM}$ is unknown, so the ratio $\Omega_A/\Omega_\text{DM}$ plays the role of a free parameter. By setting $\Omega_A=\Omega_\text{DM}$ we can obtain the lower limit on the axion mass.
The curves in Fig.~\ref{fig:omega} corresponding to different pion masses lead to virtually the same value of axion mass.
Indeed, as was shown earlier in Fig.~\ref{fig:top_results}, the results for topological susceptibility from different ensembles lie close to each other.
So, in this preliminary analysis we retain the result of Ref.~\cite{Burger:2018fvb} for the lower bound on axion mass $m_A=20(5)$~$\mu$eV.

\section{Summary}
We measured chiral observables and topological susceptibility in the region $120 \lesssim T \lesssim 600$~MeV.
The temperature dependence of $\langle\bar{\psi}\psi\rangle_3$~\eqref{eq:pbp3-def} shows clear threshold at $T\simeq 300$~MeV,
above which a trend consistent with 3D $O(4)$ scaling gives way to a simple leading order Griffith analytic behavior.
Around the same point $T\simeq 300$~MeV the topological susceptibility starts to follow  DIGA-like power-law decay.
The high-$T$ topological results from different studies are in the same ballpark, but lacking complete quantitative agreement.
Still, the final prediction for axion mass is rather insensitive to these differences. The same holds for its dependence on pion mass, once the appropriate scaling is applied.

\section*{Acknowledgments}
This work is partially supported by  STRONG-2020 under grant agreement No. 824093, RFBR grant 18-02-40126, and by the "BASIS" foundation.
Numerical simulations have been carried out on computational resources of CINECA (INFN--CINECA agreement project INF21\_sim and ISCRA project IsB20), the supercomputer of Joint Institute for Nuclear Research "Govorun", and the computing resources of the federal collective usage
center Complex for Simulation and Data Processing for Mega-science Facilities at NRC "Kurchatov Institute", http://ckp.nrcki.ru/.

\bibliographystyle{JHEP}
\bibliography{p}

\end{document}